\documentclass[sn-mathphys-num]{sn-jnl}
\usepackage[T1]{fontenc}
\usepackage[latin9]{inputenc}
\usepackage[version=3]{mhchem} 
\usepackage{graphicx}
\usepackage[dvipsnames,table]{xcolor} 
\usepackage{textcomp} 
\usepackage{lmodern} 
\usepackage{hyperref}
	
\usepackage{soul}

\begin{document}

\title{Scanning acoustic microscopy characterization of cold sprayed coatings deposited on grooved substrates}

\author*[1]{\fnm{Martin} \sur{Koller}}\email{koller@it.cas.cz}

\author[2]{\fnm{Jan} \sur{Cizek}}\email{cizek@ipp.cas.cz}

\author[1]{\fnm{Michaela} \sur{Janovsk{\'{a}}}}\email{janovska@it.cas.cz}

\author[1]{\fnm{Martin} \sur{{\v{S}}ev{\v{c}}{\'{i}}k}}\email{sevcik@it.cas.cz}

\author[3]{\fnm{Jan} \sur{Kondas}}\email{jk@impact-innovations.com}

\author[3]{\fnm{Reeti} \sur{Singh}}\email{rs@impact-innovations.com}

\author[1]{\fnm{Hanu\v{s}} \sur{Seiner}}\email{hseiner@it.cas.cz}

\affil*[1]{\orgname{Institute of Thermomechanics of the Czech Academy of Sciences}, \orgaddress{\street{Dolej\v{s}kova 1402/5}, \city{Praha~8}, \postcode{182~00}, \country{Czechia}}}

\affil[2]{\orgname{Institute of Plasma Physics of the Czech Academy of Sciences}, \orgaddress{\street{Za Slovankou 1782}, \city{Praha~8}, \postcode{182~00}, \country{Czechia}}}

\affil[3]{\orgname{Impact Innovations GmbH}, \orgaddress{\street{B\"{u}rgermeister-Steinberger-Ring~1}, \city{Rattenkirchen}, \postcode{84431}, \country{Germany}}}

\abstract{
The effect of non-planar substrate surface on homogeneity and quality of cold sprayed (CS) deposits was studied by scanning acoustic microscopy (SAM). Fe coatings were cold sprayed onto Al substrates containing artificially introduced grooves of square- and trapezoid-shaped geometries, with flat or cylindrical bottoms. The Al substrates were either wrought or cold sprayed, to comprehend their prospective influence on the Fe coatings build-up. SAM was then used to assess morphological properties of the materials from the cross-view and top-view directions. The microstructure below the surface of the studied samples was visualized by measuring the amplitudes of the reflection echoes and the velocity of the ultrasonic waves. The SAM analysis revealed that the regions of coating imperfections around the grooves are larger than what is suggested by standard scanning electron microscopy (SEM) observations. Furthermore, we found that the seemingly non-influenced coating regions that appear perfectly homogeneous and dense in SEM do, in fact, possess heterogeneous microstructure associated with the individual CS nozzle passes.}

\keywords{cold gas dynamic spray, ultrasonic characterization, additive manufacturing, machined surfaces, non-perpendicular impact}

\maketitle
\section{Introduction}

Cold spray (CS) is a modern surface coating deposition method ranked among the thermal spray technologies family. As opposed to its predecessors, it uses kinetic energy instead of thermal input for the deposition. Without melting, the individual feedstock powder particles are accelerated to very high velocities, and the coating forms through their extensive plastic deformation upon impact \cite{2011Hussain, 2016Kay, 2020Li, 2021Cizek}. That way, several detrimental effects triggered by the use of the thermal input in the other technologies are eliminated (such as oxidation or phase changes of the sprayed material), leading to coatings with improved mechanical properties (e.g., strength, hardness, adhesion) given by the inherent dense and well-connected microstructure.

As with every coating deposition method, good adhesion to a substrate or previously deposited layers is a quality of paramount importance. As recognized in the literature, the adhesive strength in CS is governed by three physical phenomena: i) mechanical anchoring, i.e., conformation to the asperities of the underlying material \cite{2018Raoelison, 2019Nikbakht, 2020Hassani}, ii) metallurgical bonding, i.e., diffusion-based binding of the two metals in intimate contact (upon breaking of their natural oxide shells triggered by the impact) \cite{2003Assadi, 2018Hassani-Gangaraj, 2019Ichikawa}, and iii) intermixing, a phenomenon observed under low deposition efficiency conditions \cite{2003Grujicic, 2005Champagne, 2018Xie, 2021Yin}, leading to larger-scale straining, fracture, and mutual embedding of the materials at the coating/substrate interface.

At some conditions, the coating/substrate interface quality may be compromised even for the CS coatings. This could happen due to a variety of reasons pertaining to poor surface preparation (excessive oxidation, presence of impurities, embedded residual grit, etc.), due to undesirable response of the materials (stress formation, deformation, or temperature-induced changes), or due to geometrical factors \cite{2018Drehmann, 2020Sample, 2018Petrackova}. The latter is frequently associated with the fact that upon spraying components of complex geometries, the impact of the individual particles may not always be perpendicular. The phenomenon of inclined-surface impact was studied previously \cite{2018Petrackova, 2007Li, 2010Binder, 2014Yin} and, as opposed to a perpendicular impact, yielded poorer microstructures (increased porosity, lower adhesion).

In this study, we perform an analysis of the heterogeneities within the CS deposits sprayed over inclined surfaces, using the scanning acoustic microscopy (SAM). In this method, ultrasonic waves are focused into a single point of a material by a concave immersion transducer, and the reflection echoes caused by discontinuities in the acoustic impedance (e.g., sample edges, phase boundaries, inner defects, etc.) are detected \cite{1974Lemons,1979Quate,1993daFonseca,1995Yu,1997Yang,2019Bertocci,2020Yu}. The SAM measurements were already used for surface coatings and layers deposited by several different techniques, and planar defects at the interface such as delamination \cite{2002Galliano, 2009Alig, 2012Oehler, 2016Bi}, crack propagation \cite{2018Schilling} or pitting corrosion \cite{2020Zhu} have been observed. Recently, the SAM technique was, for the first time, utilized also for the characterization of artificial defect repair carried out by the cold spray method \cite{2023Xu}. In that study, aluminum AA 7050 alloy was used, and the morphology of porosity around the defect (pit) was visualized at macro-scale level. Importantly, the pit was very shallow, resulting in an almost perpendicular impact of the particles. In our previous study \cite{2020Janovska}, a good mutual adhesion of an as-sprayed Fe CS coating deposited on planar Al alloy substrates was shown, and SAM was used at micro-scale level to visualize a crack that formed within an intermetallic compound present at the coating/substrate interface after a thermal cycle to 500~$^{\circ}$C.

In this paper, an SAM analysis of more complex geometry of interfaces is presented, where Fe coatings were cold sprayed onto three groove geometries machined into either CS or wrought Al. From the properties of ultrasonic waves travelling through sub-millimeter plates cut from the samples in two directions (across the sprayed grooves and parallelly to them), SAM unveils the additional locations of deteriorated material that may have a significant influence on the overall properties but cannot be directly seen by SEM observation.

\section{Materials and methods}

\subsection{Samples fabrication}
\begin{figure}[htp]
	\centering
		\includegraphics[width=1.00\textwidth]{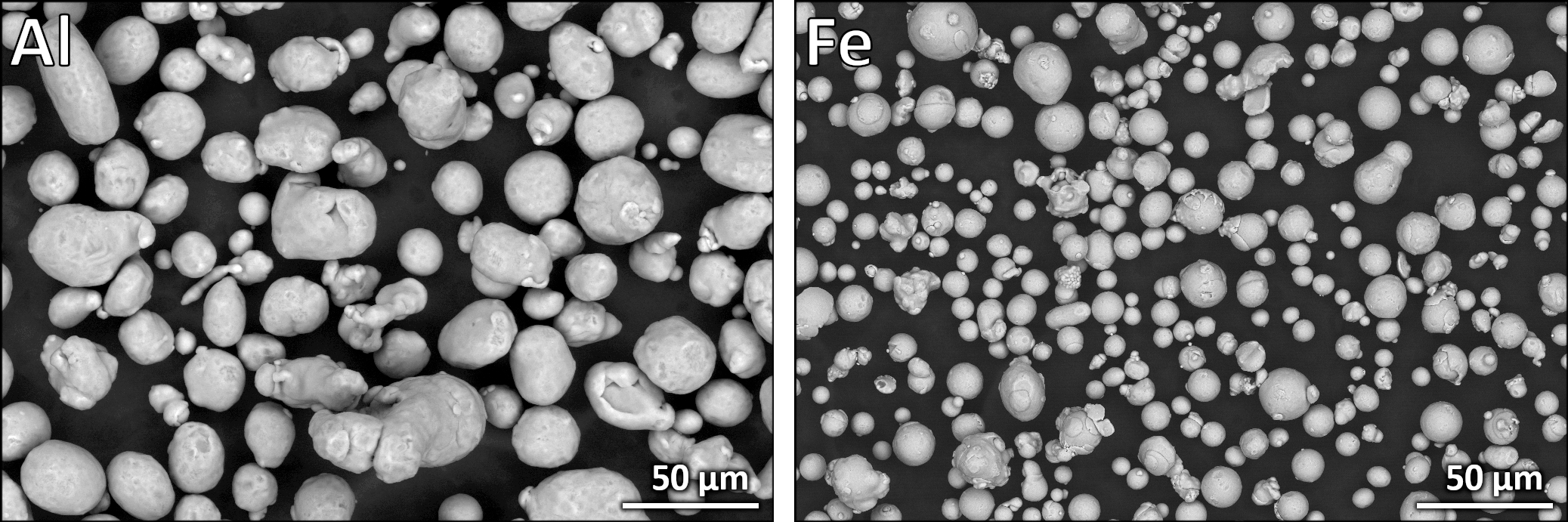}
	\caption{Morphology of the gas-atomized Al and Fe powders used for the cold spraying. Aluminum was used for the production of cold sprayed substrate (to be compared with wrought Al), while Fe was used as the final coating to overspray all grooves.}
	\label{fig:powders_AlFe}
\end{figure}

Al and Fe were used as case study materials to demonstrate the potential of the SAM method for characterization of CS deposits. Wrought or CS Al were used as the substrates into which the grooves were machined, and Fe was used as the coating overlay. The wrought aluminum was a commercially available AA~1050 alloy sheet of 99.5\% purity (EN AW-1050A H24; Alcom Alval, Czech Republic). To prevent its bending during the CS and machining processes, the thickness of the sheet was 10~mm. It was cut into 150$\times$100$\times$10~mm$^3$ blocks and was either used as a substrate directly or served as a mandrel for a CS preparation of thick Al deposits. For that, gas atomized 20--50~{\textmu}m Al powder (Toyal Europe, France) was used (Fig.~\ref{fig:powders_AlFe}), having a purity of 99.99\%.

\begin{figure}[htp]
	\centering
		\includegraphics[width=\textwidth]{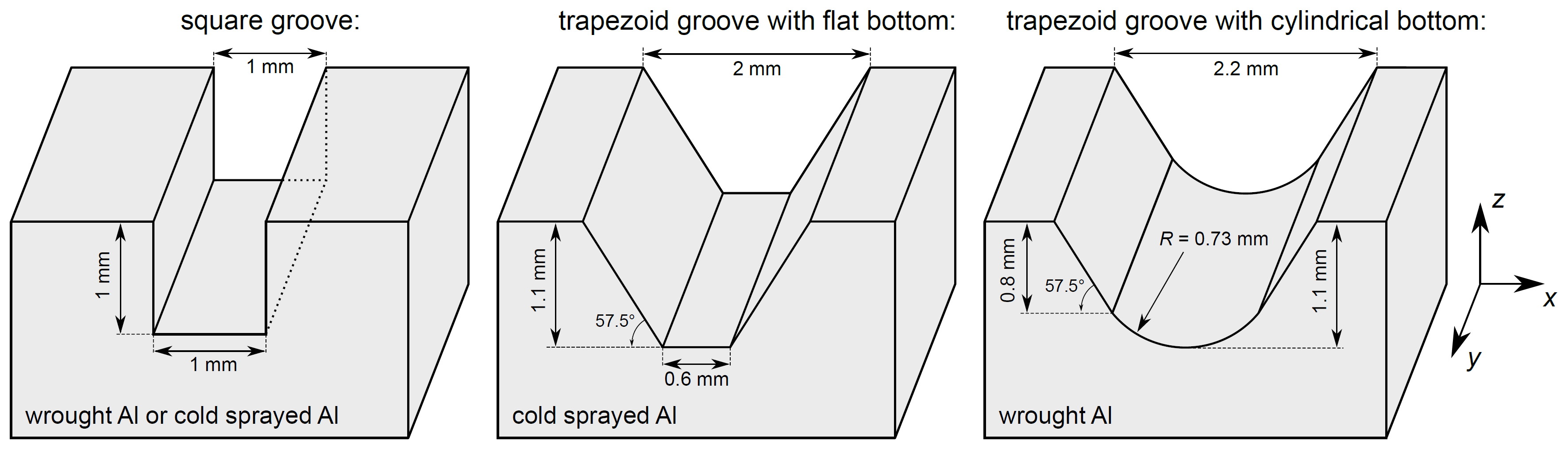}
	\caption{Geometry of the grooves machined into the Al substrates, which were subsequently cold sprayed with the Fe coating overlay. The direction $z$ corresponds to the spraying axis. Only the areas near the grooves are sketched in this figure, the used Al substrates had much larger dimensions.}
	\label{fig:grooves}
\end{figure}

Subsequently, square-shaped grooves and two types of trapezoid-shaped grooves were machined into both substrate types. In total, four different combinations with the Fe overlay were prepared. The square grooves, with dimensions of 1$\times$1~mm$^2$, were machined into both substrate types to investigate whether the slightly different material properties of the two Al materials will influence the Fe coating consolidation. The square geometry comes from the application of CS, where technological notches are made using precision CNC milling. The trapezoid grooves had lateral faces inclined at an angle of 57.5$^{\circ}$ to the surface plane. The groove machined in the CS Al substrate had a flat bottom, while the groove machined in the wrought Al substrate had a cylindrical bottom, as illustrated in Fig.~\ref{fig:grooves}. This slightly different geometry of the trapezoid grooves aimed to reveal the distinction between cases where the spraying angle changes abruptly or continuously during the spraying.
This is the case where different modes of use of CS are considered. The trapezoid with the flat bottom typically represents a situation where the grooves are original, complex segments of a component surface. Contrary to this, the cylindrical bottom represents a situation where CS is used as a repair technology and the original surface has been removed by tooling. Here, the adhesion of the repaired material is artificially enhanced by the gradual inclination \cite{2018Petrackova}.

Aside from the groove geometry, the influence of another parameter was studied, the CS torch movement direction. To do this, additional trapezoid grooves were machined in the wrought Al, this time in both principal directions $x$ and $y$ perpendicular to each other, forming a square pattern with the parallel grooves spaced 5~mm from each other (the spray direction is considered along the $z$ axis throughout this paper and the torch moved along the $x$ direction when spraying over the sample). Prior to the deposition of the Fe coatings, the Al substrate surfaces were cleaned using ethanol.

\clearpage
Thick $\geq 1.5$~mm coatings of Fe were then cold sprayed onto the grooved Al substrates. The gas-atomized Fe powder (Nanoval, Germany) particle size ranged from 10--32~{\textmu}m and its morphology is shown in Fig.~\ref{fig:powders_AlFe}. A high-pressure ISS 5/11 system (Impact Innovations, Germany) was used for the deposition of the Fe overlay coating (as well as the underlying Al). Nitrogen at 50~bar pressure and 900~$^{\circ}$C was used as the process gas (50~bar and 500~$^{\circ}$C in the case of the underlying Al) and the stand-off distance was 30~mm. The robot travel speed was 500~mm/s, and the bead distance was set to 1~mm. This setup resulted into a per-pass thickness of the Fe of approximately 270~{\textmu}m. A perpendicular spraying setup was used across the entire sample area, regardless of the actual surface inclination in the grooves.

\subsection{SAM characterization}
\begin{figure}[htp]
	\centering
		\includegraphics[width=1.0\textwidth]{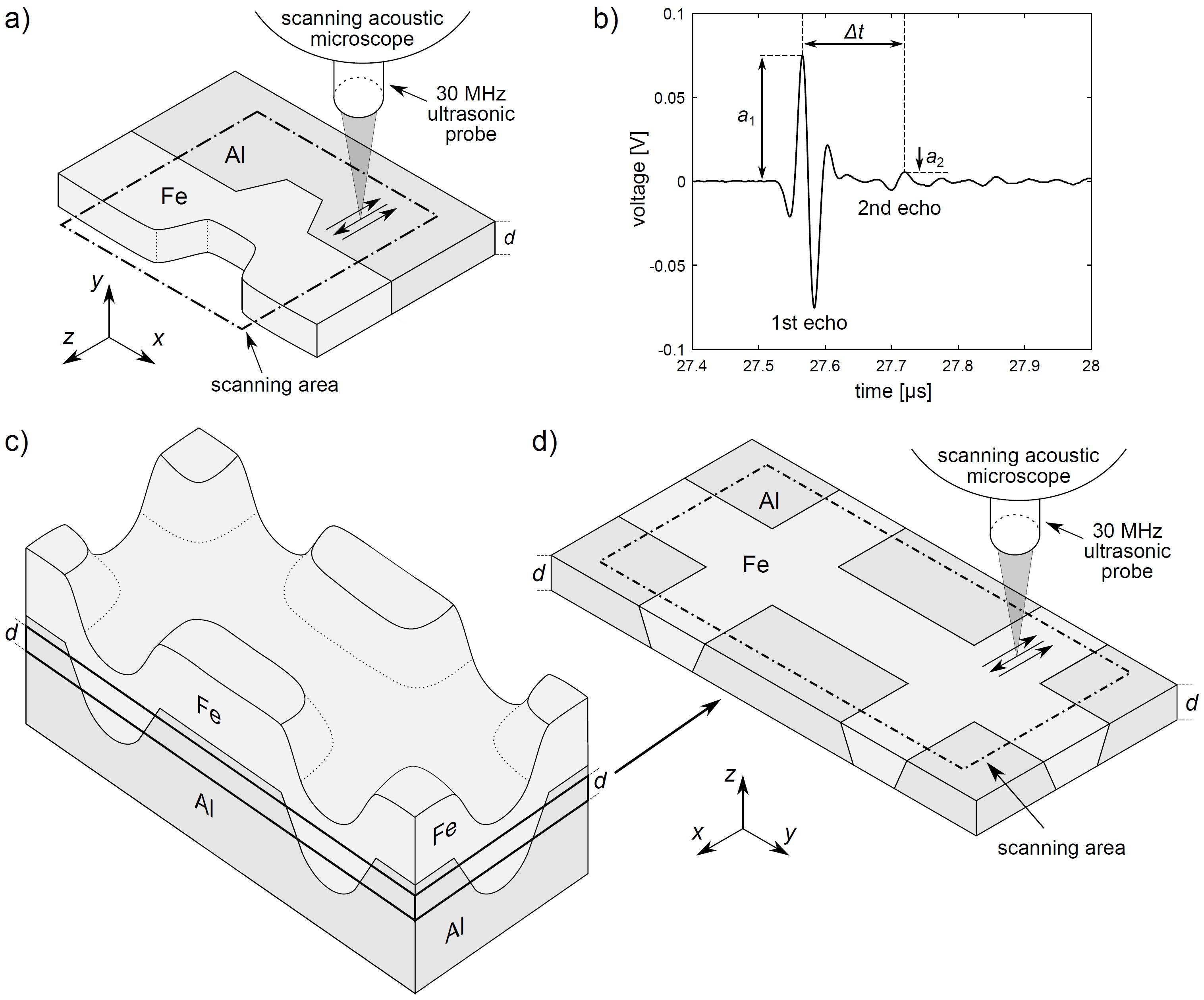}
	\caption{a) Schematics of the SAM measurements on the cross-view plates cut perpendicularly to the grooves. The scanning area comprised regions at least 1 millimeter around the grooves for all studied plates (dash-dotted regions) and the arrows at the top face highlight the scanning pattern (not to scale), b) An example of a time--voltage signal acquired at one scanning point in the Fe coating, c) Schematics of the square pattern trapezoid-grooved sample over-sprayed with Fe coating. The thick lines show the locations, from which the top-view plate was subsequently cut, d) Schematics of the SAM measurements on the top-view plate.}
	\label{fig:SAMschema}
\end{figure}

The SAM characterization was done on two different types of plates, cross-view, and top-view. For the former, a plane-parallel plate was cut across the groove from each of the four different combinations such that its thickness ($d$, varied from 0.3--0.5~mm for the samples) was parallel to the $y$ axis of the machined grooves, as shown in Fig.~\ref{fig:SAMschema}a. In this configuration, the substrate/coating interface was parallel to the axis of the transducer during the SAM measurements (i.e., to the propagation direction of the ultrasonic waves), allowing the study of the interface but also inner defects within the Fe coating or the Al substrate across the entire plate thickness. The second plate type (top-view) was cut from the sample with the square pattern of the trapezoid grooves in the wrought Al substrate. Here, the plate thickness was parallel to the spray direction $z$, and this plate contained two groove intersections, as sketched in Fig.~\ref{fig:SAMschema}d. The geometry of the top-view plate allowed studying the distribution and the extent of the coating imperfections along the entire groove length between the intersections, as well as the effect of orientation of the groove in relation to the spraying pattern. The major faces of all plates were ground by SiC \#1200 foils (Struers, Denmark) using a jig that ensured parallelism of the ground surfaces.

For the SAM experiments, Olympus UH Pulse100 scanning acoustic microscope (equipped with EA-PS 2384-05B power supply, JSR DPR500 dual pulser/receiver, high-speed NI PXI-5152 digitizer, and NI PXI-7350 motion controller) was used. The studied plates were treated with a thin protective layer of an anti-corrosive couplant, having a similar acoustic impedance to water. Each plate was then immersed in distilled water so that its thickness was parallel to the axis of the immersion transducer, as shown in Fig.~\ref{fig:SAMschema}. Here, the plates lied freely on a metallic support which ensured that, in the scanning area, both major faces were in contact only with the surrounding distilled water. Panametrics V375-SU probe was chosen as an immersion transducer, as its nominal frequency of 30~MHz was sufficient to clearly distinguish between the echoes reflected on the top and the bottom faces of the studied plates. Also, such ultrasonic waves were not fully attenuated through the sub-millimeter thickness, which allowed us to detect the signal from the ultrasonic waves reflected at the bottom face of the studied plates at the majority of the scanning points. The distance of the transducer from the studied samples was set to obtain the highest amplitude of the echo reflected at the top face. Such setup also ensured that the focal area was as small as possible. The reflection signals were then detected along a 2D scanning pattern (C-scan regime \cite{2020Yu}) from areas as far as 1~mm from the studied grooves to comprehend their potential long-range influence (Fig.~\ref{fig:SAMschema}). During the scanning route, the transducer was continuously moving with the velocity of 4~{\textmu}m/s along one direction ($z$ for the measurements on the cross-view plates, Fig.~\ref{fig:SAMschema}a, and $x$ for the top-view plate, Fig.~\ref{fig:SAMschema}d), and the reflection signals were detected at equidistant points spaced 2~{\textmu}m and 4~{\textmu}m, respectively. After the entire line was measured, the second horizontal coordinate was changed by a step of 20~{\textmu}m, and the next line was recorded with the transducer moving in the direction opposite to the previous line. 

At each scanning point, an acoustic pressure-wave packet was generated by the ultrasonic probe and sent toward the top face of the studied plate. Due to the differences in the acoustic impedance $Z = \rho \cdot v$ (where $\rho$ is the density and $v$ is the wave velocity) between the immersion medium and the studied material, the ultrasonic waves were partly reflected to the immersion medium and partly refracted into the plate. The ultrasonic waves propagated from the focal area through the plate, and upon reaching the bottom surface, they were again partly refracted and partly reflected. The reflected waves then propagated in the opposite direction, and after traveling twice through the plate thickness, they were significantly attenuated depending on the material properties of the coating or the substrate. At the top face, a certain amount of the acoustic energy was refracted back to the immersion medium, giving rise to a second echo, delayed by a few hundred nanoseconds to the first reflection echo.

In Fig.~\ref{fig:SAMschema}b, an example of the measured signal with the reflection echoes is shown, where $a_1$ represents the amplitude of the first echo (the reflection at the top face), and $a_2$ represents the amplitude of the second echo (related to the reflection at the bottom face and influenced by the wave propagation through the studied plate). The measured voltage of the received signal depended on the pulser/receiver settings. It was therefore not possible to assess the material properties in the absolute values and the relative amplitude changes were used instead. To facilitate this, the SAM settings were kept constant for all measurements. For each sample, the $a_1$ and $a_2$ amplitudes were then plotted relatively to the measurement point with the highest $a_1$ amplitude (i.e., $a_1 = 1$ at this point).

The velocity of the ultrasonic waves propagating through the studied plates was determined at each scanning point by cross-correlation between the signals of the first and the second echo. From the measured time shift values ${\Delta}t$ (Fig.~\ref{fig:SAMschema}b), the apparent wave velocity $v_{\rm app}$ was calculated as $v_{\rm app} = 2d/{\Delta}t$ (the ultrasonic waves traveled through the plate thickness twice). The term apparent is used as the obtained $v_{\rm app}$ values are an underestimation as compared to the actual wave velocities within the studied materials. This is due to the nature of the second echo: this signal also contained information from the waves that propagated through the plates not strictly along the vertical axis, but with a small deviation. Consequently, such waves traveled a distance longer than $2d$, and the measured time shift ${\Delta}t$ between the echoes was slightly increased.

To allow a direct mutual comparison of the SAM results obtained for different plates, additional ultrasonic wave velocity measurements were performed by pulse-echo (PE) method, using Olympus V208-RM probe (20~MHz nominal frequency, 3.175~mm element size). By this method, a longitudinal wave packet was sent by the PE probe through the studied sample and the signal of the echo reflected at the bottom face was detected by the same probe. In principle, both the PE method and the SAM measurements utilize the same phenomena of ultrasonic wave propagation within a material, where the wave velocities are calculated from the measured time shifts between the reflection echoes, but their main difference is in the dimensions of the affected area. In the PE method, the element size of the probe which generates the ultrasonic waves is significantly larger than the used wavelength, and thus, the generated ultrasonic waves have a planar wavefront. Therefore, the measured time shifts directly correspond to the actual wave velocity within the known sample thickness, effectively avoiding the apparent velocity issue in SAM. On the other hand, the measured PE velocity values are obtained as an average over an area of a few mm$^2$, i.e., significantly less detailed as compared to the localized SAM measurements.

The PE measurements were performed on all four cross-view plates at areas distant from the grooves. This allowed the determination of the respective longitudinal ultrasonic wave velocities for all three materials, the CS and wrought Al, and the CS Fe. The corresponding apparent wave velocity values $v_{\rm app}$ measured by the SAM were then linearly re-calibrated for each plate such that the average velocities in the Al substrate and the Fe coating matched the measured PE values. To distinguish from the apparent velocity, this re-calibrated quantity is denoted as effective velocity $v_{\rm eff}$ throughout this paper.

\subsection{SEM characterization}
After SAM, identical samples were observed using scanning electron microscopy (SEM). The SAM method typically suffices with only surface grinding for sample preparation, which is faster, but may be unusable for SEM. As such, it was necessary to use a better surface finish of the samples. At the same time, to maintain mutual comparability with the SAM results, the grinding steps were avoided and only a short polishing using 3-{\textmu}m and 1-{\textmu}m diamond pastes was carried out at low forces to prevent smearing of the soft Al material. Colloidal silica (OPS) polishing was employed as the last step. This way, also the material removal was very limited, and the surfaces observed by the SEM were identical to those previously observed using SAM. EVO MA 15 SEM (Carl Zeiss, Germany) was used at short work distances (approximately 8--10~mm) in the back-scattered mode to clearly differentiate the two metals.

\section{Results and discussion}

\subsection{Cross-view samples}

\begin{figure}[htp]
	\centering
		\includegraphics[width=0.8\textwidth]{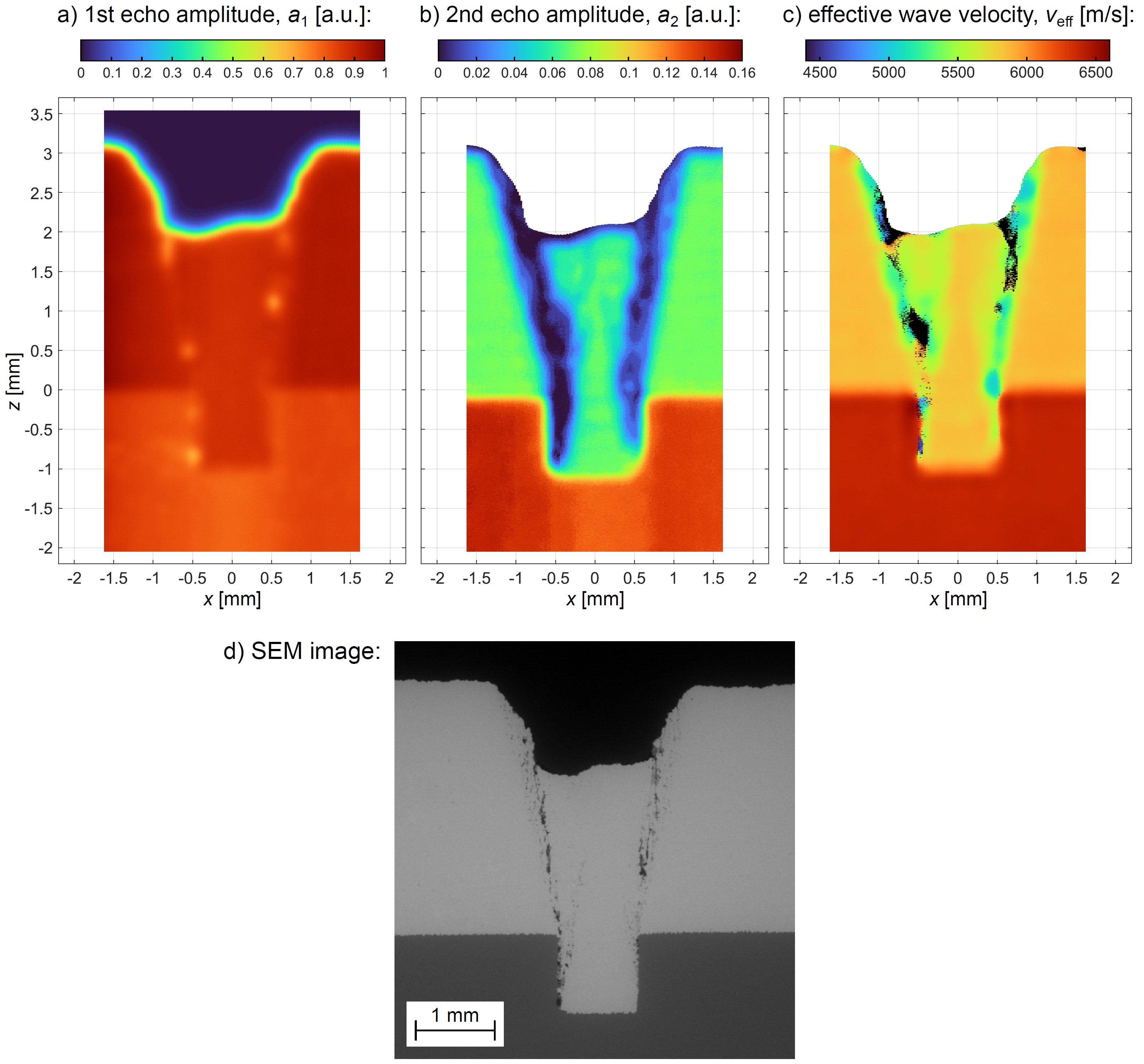}
	\caption{Sample with the square groove machined into the wrought Al substrate assessed by a)--c) SAM and d) viewed by SEM.}
	\label{fig:sq_wrought}
\end{figure}

\begin{figure}[htp]
	\centering
		\includegraphics[width=0.8\textwidth]{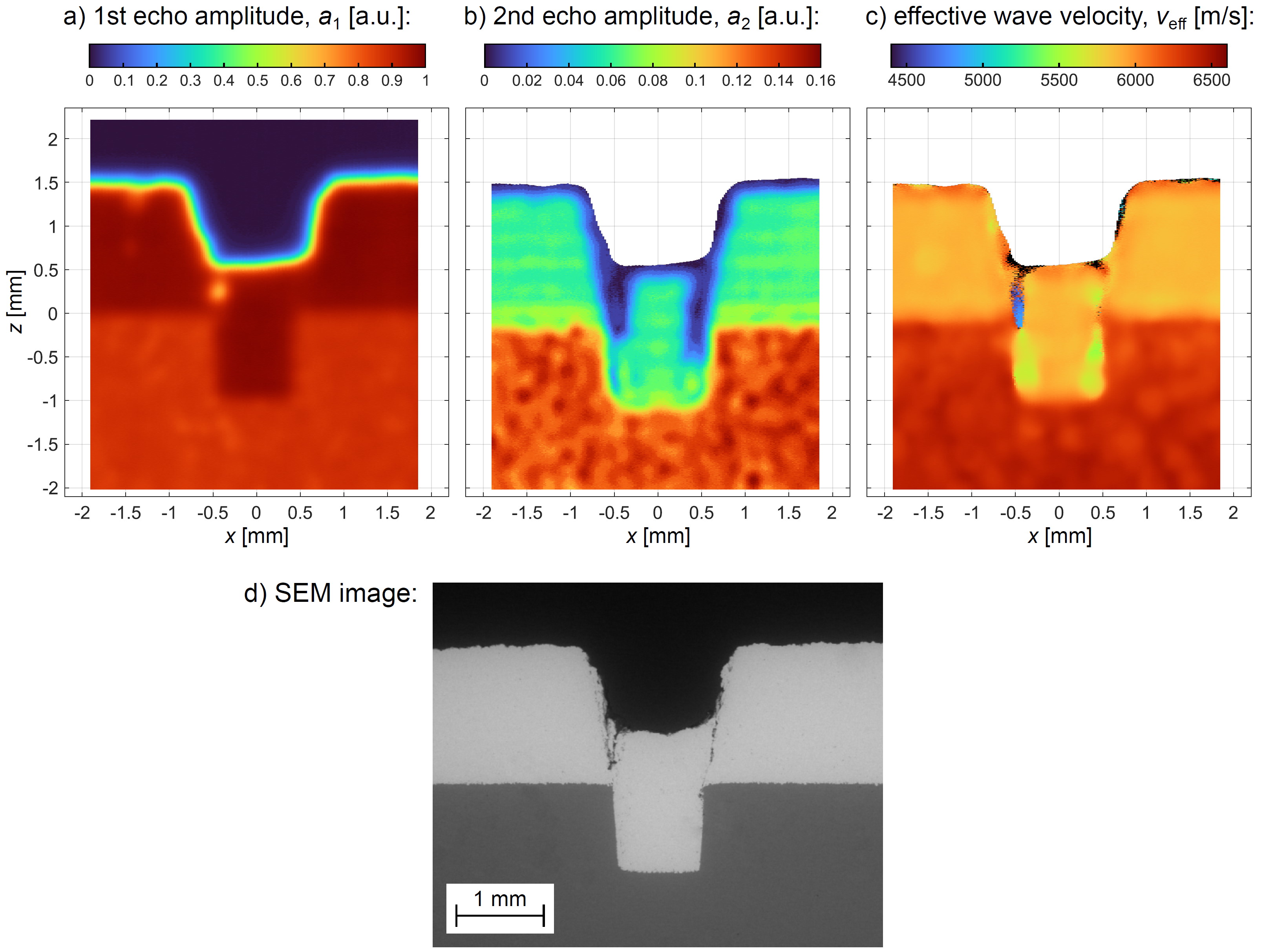}
	\caption{Sample with the square groove machined into the CS Al substrate assessed by a)--c) SAM and d) viewed by SEM.}
	\label{fig:sq_cs}
\end{figure}

\begin{figure}[htp]
	\centering
		\includegraphics[width=0.8\textwidth]{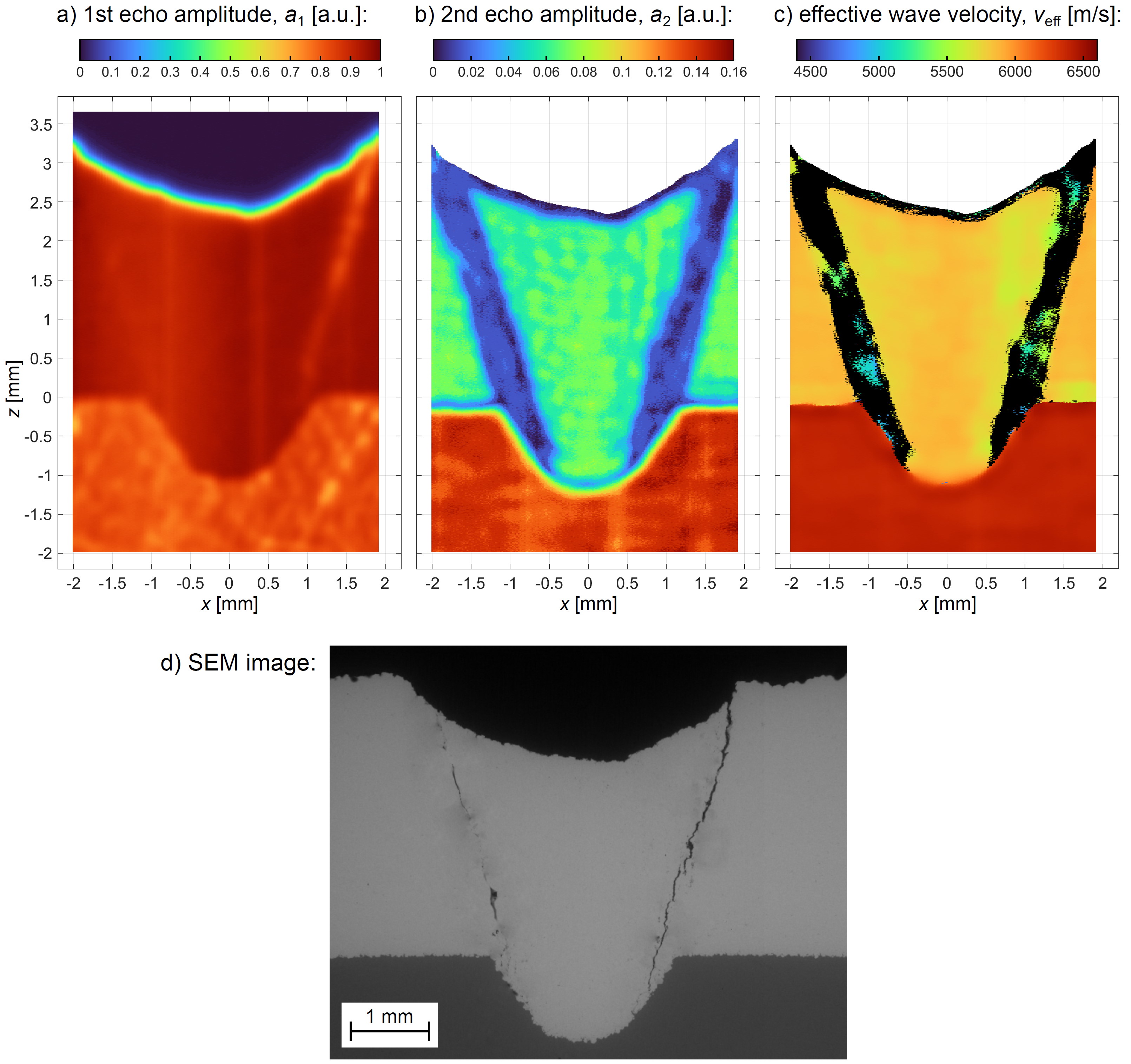}
	\caption{Sample with the trapezoid groove machined into the wrought Al substrate assessed by a)--c) SAM and d) viewed by SEM.}
	\label{fig:trap_wrought}
\end{figure}

\begin{figure}[htp]
	\centering
		\includegraphics[width=0.8\textwidth]{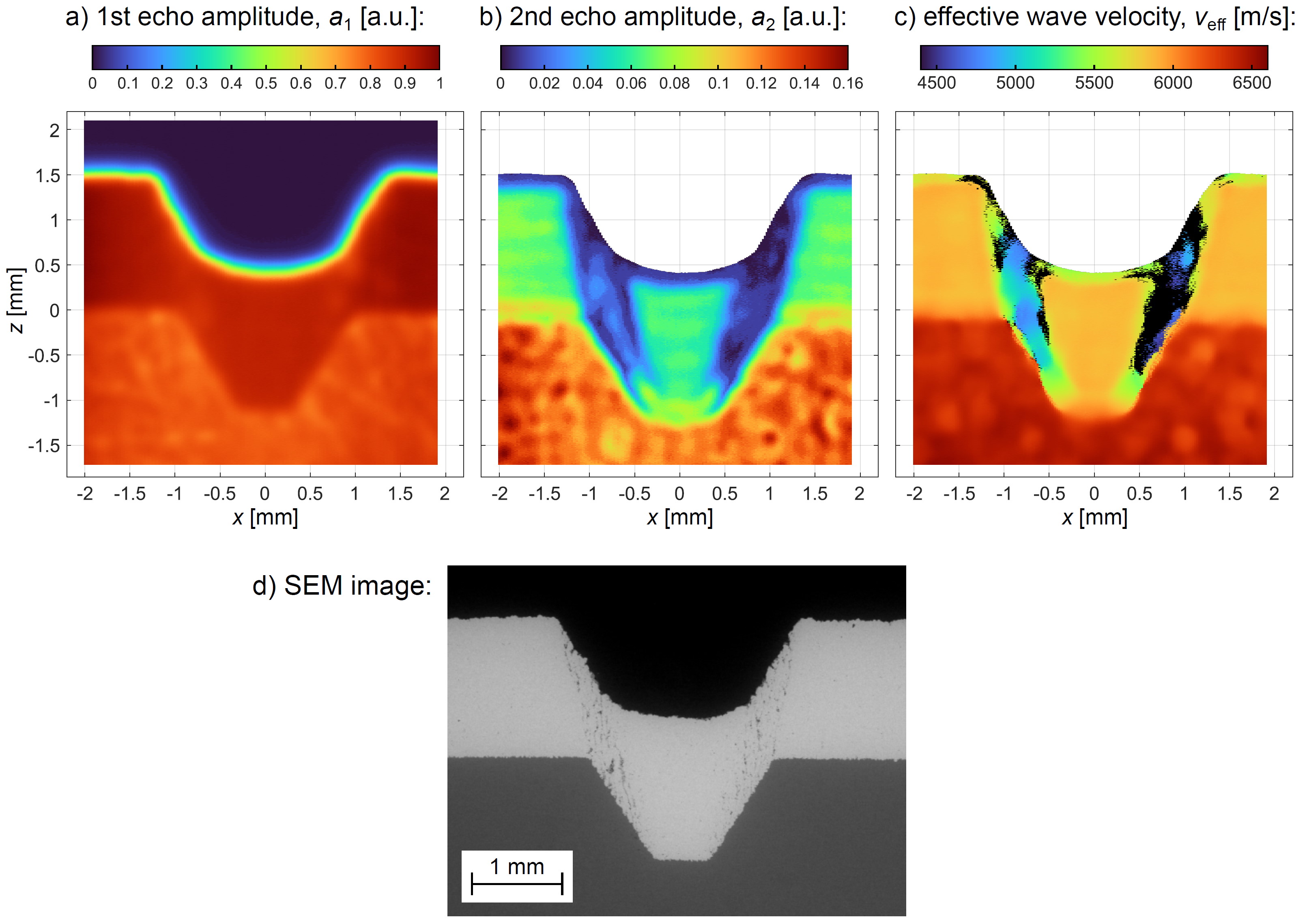}
	\caption{Sample with the trapezoid groove machined into the CS Al substrate assessed by a)--c) SAM and d) viewed by SEM.}
	\label{fig:trap_cs}
\end{figure}

The results of the SAM and SEM characterization are summarized in Figs.~\ref{fig:sq_wrought}--\ref{fig:trap_cs} for the cross-view samples. In SEM, all the CS coatings further away from the grooves appear very dense, with negligible porosity, and homogeneous throughout their entire thickness. The individual particles are not discernible, and their mutual boundaries cannot be distinguished. This testifies to the optimal selection of the high-pressure CS deposition parameters and suitability of the used powder feedstock. The interface appears faultless along the entire length of the samples, including the bottoms of all the grooves. This is somewhat surprising as the trapezoid groove with the cylindrical bottom was anticipated to generate certain levels of interface imperfections due to non-perpendicular impacts \cite{2010Binder}. 

Contrary to this, in the vicinity of the grooves, the Fe coatings exhibit notable levels of porosity in all four samples. This is a consequence of the Fe material being deposited over the groove at a moment where the coatings have already grown substantially outside the groove, having very inclined surfaces. This causes the material deposited over the groove to experience a significantly non-perpendicular impact, triggering pore formation. The porosity extent differs among the four combinations, mostly owing to the groove geometries.  In fact, in some cases, the coating imperfections can be considered macroscopic cracks rather than individual pores, such as the trapezoid groove with the cylindrical bottom, Fig.~\ref{fig:trap_wrought}d.  For this sample, the angle of incidence changed gradually during the cold spraying along the $x$ coordinate as the spray axis was always parallel to the $z$ axis. In other words, only at the center position of the groove the Fe coating was sprayed perpendicularly. The cracks observed in this sample originate symmetrically at a similar distance from the center of the groove on both sides, implying that there might be a critical angle of incidence for which the coating is not compacted when the angle is exceeded.

Regarding the ultrasonic wave propagation properties, the longitudinal wave velocities of the studied materials, determined by the PE method in the non-influenced areas distant from the grooves, are $(5871~\pm~50)$~m/s for the Fe, $(6381~\pm~50)$~m/s for the CS Al, and $(6397~\pm~50)$~m/s for the wrought Al. The measured difference in the average velocity between the two Al materials is within the experimental uncertainty. As such, the CS Al can be considered to have similar macroscopic properties to the wrought Al. Assuming the density of Al sheet as 2.70~g/cm$^3$ and the density of Fe coating as 7.81~g/cm$^3$, the longitudinal modulus $c_{11}$ can be obtained from these PE measurements as $c_{11} = \rho \cdot v^2 = 110$~GPa for the Al sheet and $c_{11} = 269$~GPa for the Fe coating. These values are in good agreement with the elasticity measurements, carried out in our previous work by resonant ultrasound spectroscopy \cite{2020Janovska}, confirming the good compaction of the CS materials outside the grooves. Note that the longitudinal modulus $c_{11}$ corresponds to the uniaxial deformation during the pressure wave propagation and, for materials with positive Poisson's ratio, it is significantly larger than the Young's modulus \cite{2020Janovska}. For Poisson's ratio of 0.34 of Al and Poisson's ratio of 0.29 of Fe, the measured values lead to Young's moduli of 72~GPa and 205~GPa, respectively.

As can be seen from the SAM $a_1$ amplitude maps in Figs.~\ref{fig:sq_wrought}a--\ref{fig:trap_cs}a, the ultrasonic wave reflectivity of the coatings and the substrates is homogenous at the locations that look dense in SEM. Some high-porosity locations within the Fe coatings observed by SEM are also visible in the $a_1$ maps, but the majority of the scanning points exceeds the $a_1$ value of 0.7, i.e., they are above 70~\% of the maximal amplitude. The coating/substrate interfaces are distinctly visible in the $a_1$ maps, as the reflectivity of the Fe coating is slightly higher than that of the Al substrates due to the higher acoustic impedance of Fe. The interfaces appear partially blurred due to two factors: the rugged interface formed during the mechanical anchoring of the Fe particles into the soft Al, and the intrinsic spot size of the SAM focal area. In conclusion, the $a_1$ maps provide similar information as SEM, as such visualizations correspond only to the surface of the studied samples.

In both the $a_2$ (Figs.~\ref{fig:sq_wrought}b--\ref{fig:trap_cs}b) and the $v_{\rm eff}$ maps (Figs.~\ref{fig:sq_wrought}c--\ref{fig:trap_cs}c), the Fe coatings are also easily distinguishable from the Al substrates. Note that in the $a_2$ maps, the 0--0.16 range is used in all graphs to maintain clarity, as the second echoes generally have much lower amplitudes than the first echoes. Regarding the substrates, there is a clear distinction between the wrought and the CS Al material in terms of homogeneity, even in areas sufficiently distant from the interfaces. In the wrought Al, the deviations in the effective velocities are ${\pm}50$~m/s (Figs.~\ref{fig:sq_wrought}c and \ref{fig:trap_wrought}c). Such values are, in fact, at the resolution limit of the used SAM setup, and thus, the wrought Al can be considered homogenous in the scale of hundreds {\textmu}m. On the other hand, in the CS Al, there appears to be a systematic distribution of areas with higher and lower $a_2$ and $v_{\rm eff}$ values, Figs.~\ref{fig:sq_cs}b,c and \ref{fig:trap_cs}b,c. Consequently, the corresponding deviations in the effective velocity $v_{\rm eff}$ in the CS Al substrate increase to ${\pm}250$~m/s. This effect originates from the CS deposition process and it will be discussed later as it stems from a similar phenomenon as in the CS Fe coating.

The material properties of the Fe coatings above the grooves look to be severely affected, as revealed by the different coloring in the $a_2$ and $v_{\rm eff}$ SAM maps. Even at several areas that look fully dense in both SEM and the $a_1$ reflectivity, the measured $v_{\rm eff}$ values can be significantly reduced, by up to 20\% in comparison to the areas of perpendicular impact. At various locations, which mostly align with the observed porosity in SEM images, the echo from the bottom surface cannot be detected at all. These locations are marked with black coloring in the corresponding $v_{\rm eff}$ maps. The propagating ultrasonic waves are often fully attenuated over a very short distance (a few hundred {\textmu}m or less) due to the wave scattering in the high-porosity areas or due to their interaction with large defects. Therefore, these SAM measurements can reveal the actual extent of the degraded material properties with more precision than SEM. This is because the SAM method reaches to a substantial depth of the studied materials instead of just showing superficial features on a selected cross-view cut.

\begin{figure}[htp]
	\centering
		\includegraphics[width=\textwidth]{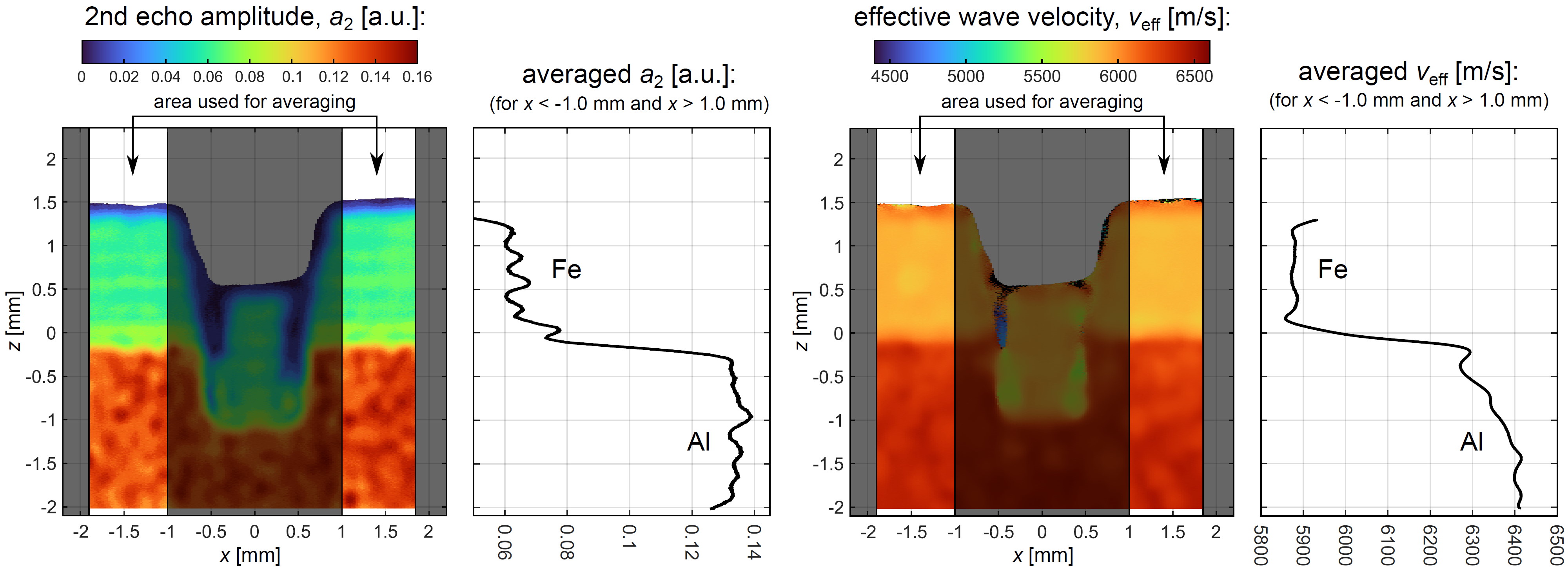}
	\caption{Periodic evolution of the wave propagation properties along the $z$ coordinates.  The unshaded areas in the SAM maps highlight the $x$ coordinates, through which the $a_2$ and $v_{\rm eff}$ values were averaged for the respective $z$ coordinates.}
	\label{fig:aver_CS}
\end{figure}

Regarding the perpendicular spraying onto planar substrates, the SAM maps surprisingly unveil intricate variations of the material properties even in areas where the Fe coating appears dense and without defects in the SEM. A somewhat systematic evolution of the amplitudes of the second echo $a_2$ is observed in the SAM maps, Figs.~\ref{fig:sq_cs}b and~\ref{fig:trap_cs}b, showing a periodic raster along the vertical $z$ coordinate in the Fe coating, independent on the $x$ coordinate. This is a feature suggesting that the CS coatings are not as homogenous, after all. To elucidate this effect, we plotted the $a_2$ and $v_{\rm eff}$ distributions averaged over the $x$ coordinates that correspond to the perpendicular spraying vs.~the $z$ coordinates for the sample with the square groove in the CS Al substrate, see Fig. \ref{fig:aver_CS}. The periodic raster is clearly seen in the $a_2$ line plot, and to some extent also in the $v_{\rm eff}$ line plot, Fig.~\ref{fig:aver_CS}. The observed oscillations have a very similar periodicity of $\sim$ 0.25--0.30~mm in the $z$ coordinate, which corresponds to the thickness of individual CS nozzle passes, approximately 270~{\textmu}m (see the experimental section). These oscillations in the Fe coating are also visible inside the grooves at the locations where the $a_2$ values are comparable to the areas distant from the groove. Moreover, similar oscillations are detected in the CS Al substrate, suggesting the observed phenomenon is not related to the used material only, but is connected to the technology instead.

For the plates with the wrought Al substrates, Figs.~\ref{fig:sq_wrought} and~\ref{fig:trap_wrought}, slight oscillations in the $a_2$ or the $v_{\rm eff}$ are also detectable in the Fe coating. However, they are less pronounced in comparison to the Fe coatings sprayed onto CS Al substrates, whose properties were influenced by the work hardening process when deposited. As such, the substrate material properties seem to have a measurable effect on the compaction of the thick CS Fe coatings, even in layers relatively far from the coating/substrate interface. 

\subsection{Top-view sample}

\begin{figure}[htp]
	\centering
		\includegraphics[width=0.75\textwidth]{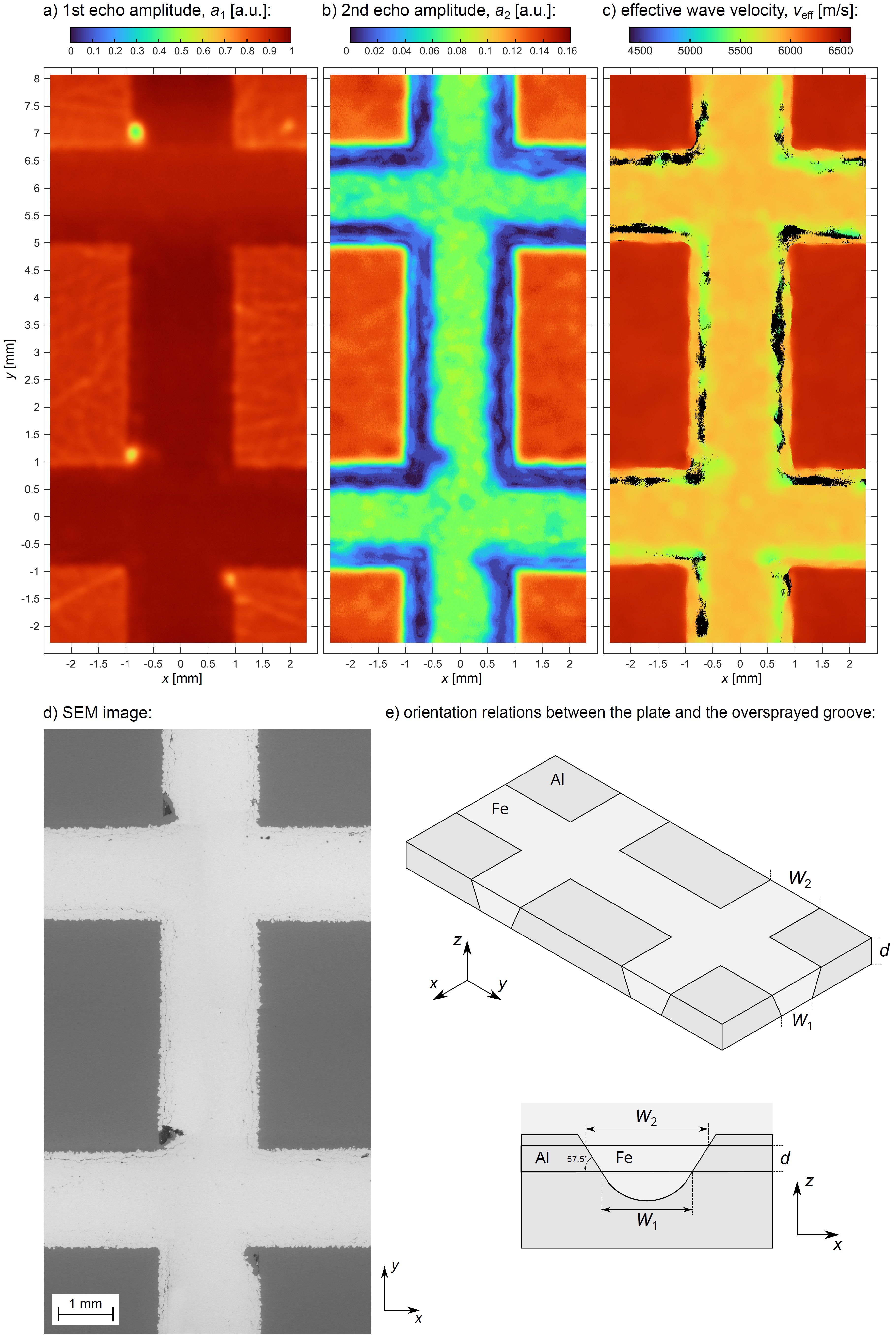}
	\caption{Top-view plate cut from the sample with the trapezoid groove in the wrought Al substrate assessed by a)--c) SAM, d) viewed by SEM, and e) orientation relations between the plate and the oversprayed groove.}
	\label{fig:FIG_topview}
\end{figure}

Figure~\ref{fig:FIG_topview} presents SAM maps measured at the top-view plate, along with the SEM micrograph. In the amplitudes of the first echo $a_1$, the boundaries between the coating and the substrate at the upper face of the plate are clearly distinguishable. The width of the Fe coating is observed to be consistent, $W_2 \approx 1.80$~mm, confirming that the top-view plate is cut perpendicularly to the spraying direction. The width of the coating at the bottom face is $W_1 \approx 1.27$~mm. The original machined trapezoid groove had a width of 2.2~mm at the top and approximately 1.18~mm at the transition point between the cylindrical bottom and the 57.5$^{\circ}$ lateral sides (Fig.~\ref{fig:grooves}). Comparing the original values with the actual geometry of the groove in the studied plate, all Fe/Al interfaces visualized by SAM in the top-view plate correspond solely to the lateral, inclined sides of the groove. In the SEM micrographs in Fig.~\ref{fig:FIG_topview}d, numerous continuous cracks are visible within the Fe coating at the upper face of the studied plate. Most of these cracks are parallel to the Fe/Al interface and their distance of 0.1--0.2~mm from the interface suggests that they are the same type of cracks as those discussed for the cross-section plate of this sample (Fig.~\ref{fig:trap_wrought}), originating from the non-perpendicular impact at the cylindrical bottom of the groove.

In the SAM maps related to the ultrasonic wave through-sample propagation, $a_2$ and $v_{\rm eff}$ in Fig.~\ref{fig:FIG_topview}, the effect of the inclined interfaces is clearly visible. When the propagating waves encounter the tilted Fe/Al interface, most of their energy is reflected or refracted in directions that no longer align with the transducer axis. As a result, the amplitude of the second echo $a_2$ is significantly reduced in the received signal, and the determined effective velocity $v_{\rm eff}$ can be also decreased or even undetectable. The previously discussed porosity in the Fe coating, resulting from the non-perpendicular impact, can also further reduce the amplitude of the second echo $a_2$. In the SAM maps in Fig.~\ref{fig:FIG_topview}a--c, the defects appear to be evenly distributed in both the $x$ and $y$ directions. Therefore, it is observed that the CS pattern does not induce any anisotropy in the $x-y$ plane, whether in relation to the perpendicular spraying on a planar substrate outside the grooves or to the non-perpendicular spraying onto tilted surfaces.

\section{Discussion of the applicability of SAM for cold spray characterization}

The presented experiments and obtained results show the potential of the SAM technique to characterize the material properties of CS deposits. For these uses of SAM, the key limitation is the penetration depth, which is influenced by several factors. First, the ultrasonic waves have to be transmitted into the studied material. The amount of acoustic wave energy transmitted across an interface depends on the differences in the acoustic impedance of the two materials forming the interface \cite{2020Yu}. This has been demonstrated by modeling studies examining the effect of various interfaces in SAM \cite{2018Marangos, 2022Shukla}. Due to the transmission effects, the SAM measurements typically cannot be performed in air or other gaseous media. Instead, water is usually used as the immersion medium as the acoustic impedances of solids are much more similar to the acoustic impedance of water than to any gaseous medium.

Another parameter influencing the SAM results is the frequency of the ultrasonic waves. At higher frequencies, the wavelength is shorter, which increases both the time resolution and the spatial resolution. Therefore, the usage of ultrasonic transducers with higher frequencies may be beneficial in the SAM measurements, as they would allow the characterization of smaller volumes of material with higher precision. On the other hand, the higher frequency usually leads to higher attenuation \cite{2020Ono}, which significantly decreases the penetration depth. Therefore, the optimal choice of the transducer and the SAM setup depends on the individual measurement. In this paper, a 30 MHz transducer was used, which was shown to provide good spatial and time resolution. This allowed us to observe not only the defect locations, but also we discovered the oscillations related to the individual spray passes that were invisible in SEM.

The SAM characterization capabilities are strongly influenced by the studied material as well. Within a polycrystalline material, the wave velocity depends on its density and elastic properties, and the attenuation of the ultrasonic wave mainly depends on internal friction and scattering, as shown in detail in \cite{2018Ryzy}. Since the various materials usually differ in these attributes, SAM measurements are applicable for distinguishing the materials within the studied samples, as shown here for the case of Fe coatings on Al substrates. Moreover, the inner pores act as additional interfaces with abrupt changes in acoustic impedance, which influence the ultrasonic wave propagation. Therefore, when the studied materials contain inner inhomogeneities, the amplitude of the second echo decreases, and the defect locations are observed by SAM.

In this work, we studied plate-shaped samples with parallel faces, where the distance between the locations where the echoes were formed was determined by their thickness $d$. Therefore, when we observed local oscillations in the $a_2$ or $v_{\rm eff}$ maps, they corresponded to local changes within the studied materials, since the distance between the echoes was constant for each sample. This approach would not be successful if the echoes were formed by reflections at unknown locations or non-straight interfaces. This limits this mode of SAM utilization for the characterization of materials in samples with complex shapes. However, given the uniform properties of the CS deposits, SAM can offer one more advantage: detection of changes in coating geometry, such as the thickness. In such a SAM analysis mode, the differences in the time of flight of the ultrasonic waves would indicate changes in the coating geometry, rather than the changes within the materials.

\section{Conclusion}

In this study, SAM was used for analyzing the quality of Fe coatings cold sprayed onto Al substrates with artificially machined grooves. From the results, the following conclusions can be drawn:
\begin{itemize}
\item The non-perpendicular impact during cold spraying resulted in an increased defect density, observable by SEM from two-dimensional cross-sections.
\item As the SAM reached below the surface by utilizing ultrasonic waves propagating through the studied materials, additional locations of degraded material properties of the CS coatings as well as the heterogeneity of its microstructure were observed. These features unveiled by SAM are not visible in the SEM micrographs.
\item The seemingly homogeneous CS coatings were observed to exhibit a periodical structure through their thickness, corresponding to the individual CS nozzle passes.
\item The macroscopic properties of the CS Al substrate were shown to be very similar to those of the wrought Al sheet, even though the CS Al is more heterogeneous at the sub-millimeter scale. Nevertheless, the slight differences of the substrates were shown to influence the consolidation of the Fe coating through its thickness.
\end{itemize}
As the SAM data contain integral information from a significant thickness of the studied material, SAM measurements can be a useful, complementary method to SEM, providing valuable insight for an inspection of cold sprayed materials in meso-scale volumes (fractions of mm$^3$ to several mm$^3$).

\backmatter
\bmhead{Acknowledgements}

This work was supported by the Czech Science Foundation project GA22-14048S and by the Operational Programme Johannes Amos Comenius of the Ministry of Education, Youth and Sport of the Czech Republic, within the frame of project Ferroic Multifunctionalities (FerrMion) [project No. CZ.02.01.01/00/22\_008/0004591], co-funded by the European Union.

\bmhead{Data$\,\,$Availability$\,\,$Statement}
The data for the SAM figures are available at https://doi.org/10.5281/zenodo.12698399.

\bibliography{Koller_JTST_rev.bib}

\end{document}